\documentclass[]{article}

\usepackage[utf8]{inputenc}   
\usepackage[T1]{fontenc}        
\usepackage{ae,aecompl}
\usepackage[english]{babel}      
\usepackage{latexsym}
\usepackage{amsmath,amsfonts}
\usepackage{epic}
\usepackage{graphicx}
\usepackage{tikz}
\usetikzlibrary{shapes.geometric}
\usetikzlibrary{decorations.pathreplacing,calligraphy}
\usepackage{calrsfs}
\DeclareMathAlphabet{\pazocal}{OMS}{zplm}{m}{n}
\usepackage{multicol}
\usepackage{hyperref} 
\usepackage{url}
\usepackage{todonotes}
\usepackage{tikz}
\usepackage{pgfplots}
\pgfplotsset{compat=1.14}
\usepackage{booktabs}
\usepackage{pgfplotstable}
\usepackage{float}
\usepackage{amssymb}
\usepackage{amsmath}
\usepackage{comment}
\usepackage{sidecap}
\usepackage{tabularx}
\usepackage{bm}
\usepackage{enumitem}
\usepackage[noend, linesnumbered,lined,boxed,commentsnumbered]{algorithm2e}

\newcommand{\mysc}[1]{\textrm{\textsc{#1}}}

\newtheorem{theorem}{Theorem}
\newtheorem{lemma}[theorem]{Lemma}

\newenvironment{proof}[1][Proof]{\begin{trivlist}
		\item[\hskip \labelsep {\bfseries #1}]}{\end{trivlist}}
\newenvironment{definition}[1][Definition]{\begin{trivlist}
		\item[\hskip \labelsep {\bfseries #1}]}{\end{trivlist}}

\newcommand{\qed}{\nobreak \ifvmode \relax \else
	\ifdim\lastskip<1.5em \hskip-\lastskip
	\hskip1.5em plus0em minus0.5em \fi \nobreak
	\vrule height0.75em width0.5em depth0.25em\fi}

\title{Special Cases of the Minimum Spanning Tree Problem under Explorable Edge and Vertex Uncertainty}

\makeatletter
\renewcommand\@date{{%
		\vspace{-\baselineskip}%
		\large\centering
		\begin{tabular}{@{}c@{}c@{}}
			Corinna Mathwieser \\
			RWTH Aachen University, Germany \\
			\normalsize \texttt{mathwieser@combi.rwth-aachen.de}
		\end{tabular}%
		\quad \quad
		\begin{tabular}{@{}c@{}c@{}}
			Eranda \c{C}ela \\
			TU Graz, Austria \\
			\normalsize \texttt{cela@math.tugraz.at}
		\end{tabular}
		
		\bigskip

		\today
}}
\makeatother

\begin{document}
	
	\maketitle
	\thispagestyle{empty}
	
	\begin{abstract}
	\noindent \emph{This article studies the Minimum Spanning Tree Problem under Explorable Uncertainty as well as a related vertex uncertainty version of the problem. We particularly consider special instance types, including cactus graphs, for which we provide randomized algorithms. We introduce the problem of finding a minimum weight spanning star under uncertainty for which we show that no algorithm can achieve constant competitive ratio.}
	\end{abstract}

\section{Introduction}
Many real world problems do not allow to work with precise data as parts of the input are uncertain or only known approximately. Different approaches to deal with uncertainty include stochastic optimization where the input data is known to follow a specific probability distribution, robust optimization which aims to find good solutions for all possible inputs and explorable uncertainty. In the latter setting, it is possible to obtain more precise or even exact data by making queries. However, any query causes exploration cost. In an applied scenario this might be time, money or other ressources which are needed for further measurements. \\
In this paper, we consider the \emph{Minimum Spanning Tree Problem under Explorable Uncertainty} (MST-U) where the goal is to find a minimum spanning tree (MST) in an edge weighted graph where the weights are initially not known but can be revealed upon request. In an instance of MST-U, each edge is equipped with an uncertainty set and a query cost. The uncertainty set, usually an interval, is guaranteed to contain the edge's weight. An edge query reveals the edge's true weight. The goal is to find a set $Q$ of queries of minimum cost which allows to find a minimum spanning tree with certainty, i.e. given the weights of all edges in $Q$, there exists an edge set $T$ such that $T$ is a minimum spanning tree for all possible weights with respect to the remaining uncertainty sets. The queries may be chosen adaptively which means that we are allowed to choose the next update based on the previous outcomes of edge queries. Note that while MST-U requires the specification of an edge set which corresponds to an MST, it is not necessary to compute the MST weight.

\subsection{Related work}
\noindent The first work on problems where parts of the input are uncertain and can be queried is due to Kahan (see \cite{kahan}) who studied the
problem of finding the maximum, the median and the minimum of a set of real values, each of which is known to lie in a given interval. Since then, explorable uncertainty has been considered for different combinatorial problems, e.g. shortest paths (see \cite{feder}), scheduling (see \cite{duerr}) and the Knapsack Problem (see \cite{goerigk}). Erlebach et al. were the first to introduce the Minimum Spanning Tree Problem under Explorable Uncertainty in \cite{erlebach}. They showed that without restrictions made on the uncertainty sets, no algorithm can achieve constant competitive ratio. This is why subsequently it is assumed that uncertainty sets are either singletons or do not contain neither their supremum nor their infimum (e.g. open intervals). \cite{erlebach} also presented the deterministic algorithm U-RED for MST-U with uniform query costs which achieves competitive ratio $2$ and proved that no deterministic algorithm can have a smaller competitive ratio. Moreover, they introduced a different version of this problem, the \emph{Minimum Spanning Tree Problem under Vertex Uncertainty} (V-MST-U) where vertices are points with uncertain locations in the plane and the edge weights correspond to the distances between the respective end vertices. They showed that U-RED can be adapted to work for the vertex uncertainty setting as well if uncertainty sets are (topologically) open. An important question left open was the effect of randomization in the setting of MST-U. This question was subsequently answered by Megow et al. in \cite{megow} where they provided the randomized algorithm \mysc{Random} with competitive ratio $1+\frac{1}{\sqrt{2}}$. The best known bound for the performance of randomized algorithms is $1.5$ which holds true even for triangles and was observed by Erlebach and Hoffmann in \cite{erlebachhoffmann}. \cite{megow} transformed their randomized algorithm into the deterministic algorithm \mysc{Balance} which achieves a competitive ratio of $2$ on instances with general query cost. \cite{megow} also considered the problem of finding the weight of an MST under uncertainty, the problem of finding an $\alpha$-approximate MST under uncertainty as well as a version of MST-U where queries return subintervals instead of the precise edge weights. Test results for MST-U are provided by Focke et al. in \cite{focke}. In \cite{verification}, Erlebach and Hoffmann deal with the \emph{verification problem} for MST-U, i.e. the problem of computing an optimal query set if the uncertainty sets as well as the exact edge weights are given. They show that the verification problem for MST-U is solvable in polynomial time while the verification problem for the vertex uncertainty problem V-MST-U is NP-hard.

\subsection{Our Contribution}

\noindent One of the main difficulties in solving MST-U stems from dealing with edges that are part of several cycles. Thus, it is a natural question to consider the special case of cactus graphs where every edge belongs to at most one cycle. However, the algorithm \mysc{Random} presented in \cite{megow} reaches its worst case competitive ratio of $1+\frac{1}{\sqrt{2}} \approx 1.71$ even for instances where the input graph is a cycle. We introduce an algorithm which achieves competitive ratio $1.5$ on instances with cactus graphs. V-MST-U on the contrary, has only been considered in the setting of deterministic algorithms so far. We prove that there exists no randomized algorithm for V-MST-U with a competitive ratio better than $2.5$ even if the input graph is a cycle. Unfortunately, structural differences between the edge uncertainty setting and the vertex uncertainty setting impede the straight forward adaption of \mysc{Random} to the vertex uncertainty setting. Instead, we consider the special case of cactus-like instances where no two cycles share a non-trivial vertex and introduce the algorithm V-\mysc{Random}$_C$ for which we prove a competitive ratio of at most $2.5$. While deterministic and randomized algorithms with reasonable performance guarantee exist for MST-U, this is not necessarily the case when aiming for more restricted graph classes than spanning trees. We demonstrate this by introducing the \emph{Minimum Spanning Star Problem under Explorable Uncertainty} (MSS-U). MSS-U is defined analogously to MST-U except that we want to identify a spanning star of minimum weight rather than a general spanning tree. For MSS-U we derive a negative result with respect to competitive analysis, i.e we show that no algorithm for MSS-U can achieve constant competitive ratio.\\
\noindent The remainder of this paper is organized as follows: In Section 2 we provide the precise definition of MST-U and further basic definitions and concepts used throughout the rest of the paper. Section 3 deals with a  randomized algorithm for MST-U on cactus graphs. In Section 4 we prove a lower bound for the performance of randomized algorithms for V-MST-U and introduce an optimal randomized algorithm for V-MST-U on cactus-like graphs. In Section 5 we study the Minimum Spanning Star Problem under Explorable Uncertainty. We conclude with a brief summary and some open questions in Section 6.

\section{Preliminaries}
In this section we will introduce definitions, notation and structural results used throughout this paper. 
\subsection{Problem definition MST-U and notation}
\begin{definition}
	An \emph{(edge-)uncertainty graph} is a tuple $\mathcal{G}=(G,(A_e)_e)$ where  $G = (V, E)$ is an undirected, connected graph and for each edge $e \in E$, $A_e \subset \mathbb{R}$  is either a singleton set or a set which neither contains its infimum nor its supremum. The sets $A_e$ with $e\in E$ are called \emph{uncertainty sets}.
\end{definition}
An instance of the Minimum Spanning Tree Problem under Explorable Uncertainty (MST-U) is specified in terms of an uncertainty graph $\mathcal{G}=(G,(A_e)_e)$ as well as an a priori unknown edge weight $w_e \in A_e$ and a query cost $q_e>0$ for each $e\in E$. We denote by $n=\vert V \vert$ the number of vertices and by $m=\vert E\vert$ the number of edges of $G$. If $A_e$ contains a single element only we say that $A_e$ is \emph{trivial}. An edge $e$ is trivial if $A_e$ is trivial. For an edge $e \in E$, we denote by $L_e:=\inf A_e$ the infimum and by $U_e := \sup A_e$ the supremum of the uncertainty set. We will also refer to $L_e$ and $U_e$ as \emph{the lower and the upper limit} of $A_e$ (or of $e$) respectively. We can query an edge $e$ to determine its weight $w_e$. If we query $e$, the set $A_e$ is updated to a singleton set containing only $w_e$. The cost of querying an edge $e$ is $q_e >0$. The goal is to find an edge set $T$ of a minimum spanning tree (MST) in $G$ with respect to edge weights $w_e$ while minimizing the total cost of queries needed to find $T$. More precisely, a feasible query set is defined as follows:

\begin{definition}
	Given an uncertainty graph $\mathcal{G}$ with graph $G=(V,E)$ and uncertainty sets $A_e$, $e \in E$, as well as edge weights $w_e \in A_e$ for $e\in E$, a set $Q\subset E$ is called a \emph{feasible query set} if there exists a spanning tree $T$ in $G$ such that $T$ has minimum weight with respect to any weight function $\bar{w}:E\rightarrow \mathbb{R}$ which fulfills that $\bar{w}(e)=w_e$ if $e \in Q$ and $\bar{w}(e) \in A_e$ if $e \in E-Q$. We say that $Q$ \emph{verifies} $T$.
\end{definition}	 
MST-U thus consists in finding a feasible query set $Q$ of minimum query cost $\sum_{e \in Q} q_e$.\\
\noindent  Moreover, we will use the following definition: Given an uncertainty graph $\mathcal{G}=(G,(A_e)_e)$ and a cycle $C$ in $G$, we say that an edge $f$ is \emph{always maximal} in $C$ if $L_f \geq U_e$ for all $e$ in $C-f$.

\subsection{Vertex uncertainty problem}

In an instance of the Minimum Spanning Tree Problem under Explorable Vertex Uncertainty (V-MST-U) we are given an undirected, connected graph $G$ where each vertex corresponds to a point in the Euclidean plane. The weight of an edge is determined by the distance between its end vertices. Instead of the precise location of a vertex $v$ we are given an uncertainty set $A_v \subset \mathbb{R}^2$. $A_v$ can either be a singleton set (in which case we refer to the vertex and the uncertainty set as trivial) or an open subset of $\mathbb{R}^2$. An algorithm can query a vertex $v$ at query cost $q_v$ to reveal its exact location. We define a \emph{(vertex-)uncertainty graph} as well as a \emph{feasible (vertex) query set} in the same way as in the setting of MST-U. Then V-MST-U is defined analogously to MST-U, i.e. we want to identify a feasible vertex query set of minimum query cost. \\
\noindent In order to apply methods developed for MST-U to V-MST-U it is common to transform vertex uncertainty sets into edge uncertainty sets by computing all possible distances between vertices. The resulting instance is referred to as associated edge instance: 

\begin{definition}
	Given an instance $\mathcal{I}$ of V-MST-U with graph $G=(V,E)$ and uncertainty sets $A_v$, $v \in V$, the \emph{associated edge instance} $\mathcal{I}'$ is an instance of MST-U with graph $G$ and uncertainty sets $A_{\{u,v\}} = \{d(u',v') \vert u' \in A_u, v' \in A_v\}$.
\end{definition}
\subsection{Performance analysis}
To analyze the quality of a solution found by an algorithm we compute the competitive ratio between the query cost of the algorithm's solution and the cost of an optimal query set. An optimal query set is an optimal solution to the offline problem where all edge weights (or vertex positions) are known a priori.

\begin{definition} \label{performance_def}
	Let $\mathcal{I}$ be an instance of MST-U (or V-MST-U). By $OPT(\mathcal{I})$ we denote the cost of an optimal query set for instance $\mathcal{I}$. For an algorithm $ALG$ we denote by $ALG(\mathcal{I})$ the cost of the query set which the algorithm outputs when applied to $\mathcal{I}$. We say that $ALG$ achieves \emph{competitive ratio} $c \geq 1$ or is $c$\emph{-competitive} if 
	\begin{equation*}
		\frac{ALG(\mathcal{I})}{OPT(\mathcal{I})} \leq c
	\end{equation*} for all instances $\mathcal{I}$.
	A randomized algorithm is said to achieve competitive ratio $c \geq 1$ if the ratio between the expected query cost of the algorithm's solution and the cost of an optimal solution is at most $c$, i.e. if 
	\begin{equation*}
		\frac{\mathbb{E}(ALG(\mathcal{I}))}{OPT(\mathcal{I})} \leq c
	\end{equation*} for all instances $\mathcal{I}$.
\end{definition}

\subsection{Structural aspects of MST-U and V-MST-U}

In the following we will recall some structural insights into MST-U which were provided by Megow et al in \cite{megow} and will be used throughout the remainder of the paper. Given an instance of MST-U, a \emph{lower limit tree} is the edge set of a minimum spanning tree in $G$ where all edge weights are set equal to the lower limit of the edge's uncertainty set. An \emph{upper limit tree} is defined analogously. 

\begin{lemma} \label{preproc}
	(\cite{megow}) Let $T_L,$ $T_U$ be a lower and an upper limit tree respectively.  All edges in $T_L \backslash T_U$ with non-trivial uncertainty sets lie in any feasible query set. 
\end{lemma}

\noindent Hence, the instance can be preprocessed by querying all edges in $T_L \backslash T_U$ and we may thus assume that $T_L=T_U$.  Lemma \ref{preproc} can be translated to the setting of vertex uncertainty in the following:

\begin{lemma} \label{preproc_v}
	Given a V-MST-U instance $\mathcal{I}$, let $T_L,$ $T_U$ be a lower and an upper limit tree of the associated edge instance $\mathcal{I}'$. Then every feasible query set for $\mathcal{I}$ contains a vertex cover of $\{e \in T_L \backslash T_U \vert$e contains a non-trivial vertex$\}$ which consists of non-trivial vertices only. 
\end{lemma}

\noindent Now recall the following well-known properties of minimum spanning trees: Given a cycle $C$ in a weighted graph $G$, we have that if an edge $e$ is such that $w_e > w_{e'}$ for all $e' \in C-e$ then $e$ is not contained in any MST. If $w_e \geq w_{e'}$ for all $e' \in C$ then there exists an MST of $G$ which does not contain $e$. In the context of MST-U this means that if we encounter a cyle $C$ with an always maximal edge $e$ then $e$ can be discarded in the search for an MST. Conversely, given a cycle $C$ and a non-trivial edge $f$ which has largest upper limit $U_f$ in $C$, it is impossible to verify an MST which contains $f$ without querying $f$ because $f$ is a candidate for an edge with strictly largest weight in $C$. Possible candidates for a largest weight edge in $C$ are $f$ and all edges $e$ in $C$ with possibly larger weight than $f$, i.e. with $U_e > L_f$. We will refer to these edges as \emph{neighbors} of $f$ and denote the set of all neighbors by $X(f):=\{e\in C-f \vert U_e > L_f\}$. \cite{megow} prove the following with respect to $f$ and its neighbor set:

\begin{lemma} \label{lem2}
	(\cite{megow}) Let $T_U$ be an upper limit tree. Let $f$ be an edge in $G-T_U$ and let $C$ be the cycle in $T_U+f$. If no edge in $C$ is always maximal then any feasible query set contains $f$ or $X(f)$. Moreover, if there is an edge $g\neq f$ in $C$ such that $L_g \geq L_f$ and $U_g \neq L_f$ then $f$ lies in any feasible query set. \\
	If $T_U=T_L$ is a lower limit tree too and no edge is known to have maximum weight even after querying $f$ or all edges in $X(f)$ then an edge $g$ in $C$ with maximum upper limit $U_g$ lies in any feasible query set.
\end{lemma}

\noindent In the context of V-MST-U, we can derive the following result from Lemma \ref{lem2}: 
\begin{lemma} \label{lem2_v}
	Consider an instance $\mathcal{I}$ of V-MST-U and let $T_U$ be an upper limit tree for the associated edge instance $\mathcal{I}'$. Let $f$ be an edge in $G-T_U$ and let $C$ be the cycle in $T_U+f$. If no edge in $C$ is always maximal in $\mathcal{I}'$ then any feasible query set for $\mathcal{I}$ contains a non-trivial vertex in $f$ or a vertex cover of $\{e \in X(f)\vert$e contains a non-trivial vertex$\}$ which consists of non-trivial vertices only. Moreover, if there is an edge $g\neq f$ in $C$ such that $L_g \geq L_f$ and $U_g \neq L_f$ then any feasible query set of $\mathcal{I}$ contains a non-trivial vertex in $f$.
\end{lemma}   

\noindent We omit a formal proof of Lemma \ref{preproc_v} and of Lemma \ref{lem2_v} as they follow directly from Lemma \ref{preproc} and Lemma \ref{lem2} and the fact that an MST for a vertex-uncertainty graph can only be identified if the edge-uncertainty graph of the associated edge instance allows to identify an MST. 

\section{A randomized algorithm for MST-U on cactus graphs}

In this section we consider a randomized algorithm for MST-U on cactus graphs. A \emph{cactus graph} is a connected graph in which any two cycles share at most one vertex. A feasible query set needs to allow for the detection of a maximum weight edge in each cycle. It can be easily seen that an optimal query set for a cactus graph consists of the disjoint union of the optimal query sets for each of the graph's cycles. Thus we can first consider MST-U on a cycle and then extend the result to the case of an edge uncertainty graph which is a cactus. Once we are able to treat cycles separately, it is possible to achieve an optimal competitive ratio of $1.5$ using the following observation: Assume that we have preprocessed the instance such that $T_L=T_U$ and let $f$ be an edge in $G-T_L$ and let $C$ be the cycle in $T_L+f$. Assume moreover that no edge in $C$ is always maximal. Then Lemma \ref{lem2} guarantees that if we start by querying $f$ and continue to query edges in order of decreasing upper limit until an edge in $C$ is always maximal, then we have queried at most one edge which is not in the optimal solution.

\begin{theorem}\label{qualityrandomc}
	For	cactus graphs there exists an algorithm \mysc{Random}$_C$ with competitive ratio at most $1.5$, which is optimal. Moreover, if $G$ is a cycle, \mysc{Random}$_C$ achieves competitive ratio $1+\frac{q_{X(f)}\cdot q_f}{q_{X(f)}^2 +q_f^2}$ where $q_{X(f)}$ denotes the cumulative query cost of edges in $X(f)$.
\end{theorem} 

\begin{proof}
	We first consider a graph $C$ which consists of a single cycle. Assume again that we have preprocessed the instance such that $T_L=T_U$ and that no always maximal edge in $C$ is known. Let $f$ be the edge in $E \setminus T_L$ and let $q_{X(f)}:=\sum_{e\in X(f_i)}q_e$ be the query cost of all neighbors of $f$ in $C$. With probability $p$, our algorithm starts by querying all edges in $X(f)$. With probability $1-p$, its first step is to query $f$. Once it has queried $X(f)$ or $f$, it queries edges in order of decreasing upper limit until an always maximal edge in $C$ can be identified.\\
	We distinguish two cases: either an optimal solution queries $f$ or it does not. If an optimal solution does not query $f$, it must by Lemma \ref{lem2} query all of the neighbors in $X(f)$ and thus makes queries at cost $q_{X(f)}$. In this case with probability $p$, we query the same edges as the optimal solution and achieve competitive ratio $1$. With probability $1-p$, \mysc{Random}$_C$ queries $f$ and possibly all edges in the neighbor set $X(f)$ such that the competitive ratio is at most $\frac{q_{X(f)}+q_f}{q_{X(f)}}$. Thus, in this case the overall competitive ratio is at most $1+
	(1-p)\frac{q_f}{q_{X(f)}}$.\\
	If an optimal solution queries $f$, \mysc{Random}$_C$ queries the same edges as the optimal solution if it starts by querying $f$, i.e. with probability $1-p$. With probability $p$, \mysc{Random}$_C$ starts by querying all neighbors in $X(f)$ and might have to query $f$ too, while an optimal query set might contain $f$ only. Summing up, the competitive ratio is bounded by $1+p\cdot\frac{q_{X(f)}}{q_f}$ in this case. By setting $p=\frac{q_{f}^2}{q_{f}^2 + q_{X(f)}^2}$, the obtained bounds for both cases coincide and equal
	\begin{equation*}
		1+ \frac{q_{X(f)}  q_{f}}{q_{f}^2 + q_{X(f)}^2}  . 
	\end{equation*}
	
	\noindent Note that the following equivalences hold:
	\begin{alignat*}{2}
		\frac{q_{X(f)} \cdot q_{f}}{q_{f}^2 + q_{X(f)}^2} &\leq \frac{1}{2} &&\Leftrightarrow \\
		2q_{X(f)} q_{f} &\leq q_{f}^2 + q_{X(f)}^2 &&\Leftrightarrow \\
		0 &\leq (q_{f}- q_{X(f)})^2. 
	\end{alignat*}
	Thus, \mysc{Random}$_C$ is 1.5-competitive on instances where the uncertainty graph is a cycle if $p$ is chosen as above. \\
	For an instance $\mathcal{I}$ where the uncertainty graph is a general cactus $G$, we again preprocess the instance such that $T_L=T_U$ is a lower and an upper limit tree. Denote the edges in $G-T_L$ by $f_1,...,f_{m-n+1}$, the order is arbitrary. Let $C_i$ denote the cycle in $T_L+f_i$. We apply \mysc{Random}$_C$ for cycles as described above to each of the cycles seperately. Let $Q^*_i$ denote an optimal query set for an instance where the uncertainty graph consists only of $C_i$ and the uncertainty sets and weights are as in $\mathcal{I}$. We set $OPT_i:= \vert Q^*_i \vert$, $i=1,...,m-n+1$. As any two cycles in $G$ do not share common edges and edges that lie in no cycle need to be part of any spanning tree, the disjoint union $Q^*:=\dot\bigcup_{i=1}^{m-n+1} Q^*_i$ is an optimal solution for $\mathcal{I}$ and thus $OPT(\mathcal{I})= \sum_{i=1}^{m-n+1}OPT_i$. Moreover, the structure of cactus graphs guarantees that for any $i=1,...,m-n+1$, $X(f_i)$ is independent of the choice of queries that \mysc{Random}$_C$ makes when applied to $C_j$ with $j \neq i$ as well as the queries' outcome. Let $q_{prep}$ denote the cost of queries made in the preprocessing. We denote by $ALG_i$ the cost of the queries \mysc{Random}$_C$ makes when applied to $C_i$, $i\in \{1,2,...,m-n+1\}$. As we expect at most $1.5 \cdot OPT_i$ queries when applying \mysc{Random}$_C$ to $C_i$, we obtain that 
	\begin{align*}
		\frac{\mathbb{E}[\mysc{Random}_C(\mathcal{I})]}{OPT(\mathcal{I})}&= \frac{q_{prep}+ \sum_{i=1}^{m-n+1} \mathbb{E}[ALG_i(\mathcal{I})]}{OPT(\mathcal{I})} \\&\leq \frac{q_{prep} +\sum_{i=1}^{m-n+1} 1.5 \cdot OPT_i}{q_{prep}+ \sum_{i=1}^{m-n+1} OPT_i} \\&\leq 1.5.  
	\end{align*}
	This completes the proof of the competitiveness. Note that the correctness of \mysc{Random}$_C$ is immediate as for each cycle the algorithm proceeds to query edges until an edge becomes always maximal. The optimality follows from the fact that no randomized algorithm can achieve a competitive ratio less than 1.5 even on triangles, see \cite{erlebachhoffmann}. A precise description of \mysc{Random}$_C$ for general query costs is given in Algorithm \ref{generalrandom_c}.
	
\end{proof}

\IncMargin{1em}
\begin{algorithm}[ht]
	\SetKwData{Left}{left}\SetKwData{This}{this}\SetKwData{Up}{up}
	\SetKwFunction{Union}{Union}\SetKwFunction{FindCompress}{FindCompress}
	\SetKwInOut{Input}{input}\SetKwInOut{Output}{output}
	\Input{An instance of MST-U with cactus graph $G=(V,E)$, uncertainty sets $A_e$ and query costs $q_e$, $e \in E$}
	\Output{A feasible query set $Q$}
	\BlankLine
	Draw $b$ uniformly at random from $[0,1]$\;
	Preprocess the instance such that $T_L=T_U$\;
	Index the edges $f_1, . . . , f_{m-n+1}$ in $R := E \setminus T_L$ arbitrarily \;
	Initialize $Q=\emptyset$\;
	\For{$i\leftarrow 1$ \KwTo $m-n+1$}{
		Add $f_i$ to $T_L$ and let $C_i$ be the unique cycle closed\;
		Let $X(f_i)$ be the set of edges $g \in T_L \cap C_i$ with $U_g > L_{f_i}$\;
		Let $q_i:= \sum_{e\in X(f_i)}q_e$ \;
		
		\If{$X(f_i) \neq \emptyset$}{
			\If{$b \leq \frac{q_{f_i}^2}{q_{f_i}^2 + q_i^2}$}{add all edges in $X(f_i)$ to $Q$ and query them.}
			\Else{Add $f_i$ to $Q$ and query $f_i$.}
			\While{no edge in the cycle $C_i$ is always maximal}{Query an unqueried edge $e \in C_i \setminus Q$ with maximum $U_e$ and add it to $Q$.}}}
	\caption{The algorithm \mysc{Random}$_C$ for MST-U with general query costs in cactus graphs}\label{generalrandom_c}
\end{algorithm}\DecMargin{1em}

\section{V-MST-U and randomization}

We will now turn to the vertex uncertainty version of MST-U which has not been considered in the context of randomized algorithms so far. We start by showing that no randomized algorithm for V-MST-U can achieve a better competitive ratio than $2.5$. Our proof is based on the proof of the bound for the deterministic performance guarantee by \cite{erlebach}. 

\begin{theorem}\label{vertexbound_d}
	No randomized algorithm for V-MST-U can achieve a competitive ratio less than $2.5$. This remains true even for cycles and under the assumption of uniform query costs.
\end{theorem}

\begin{figure}
	\centering	
	\input{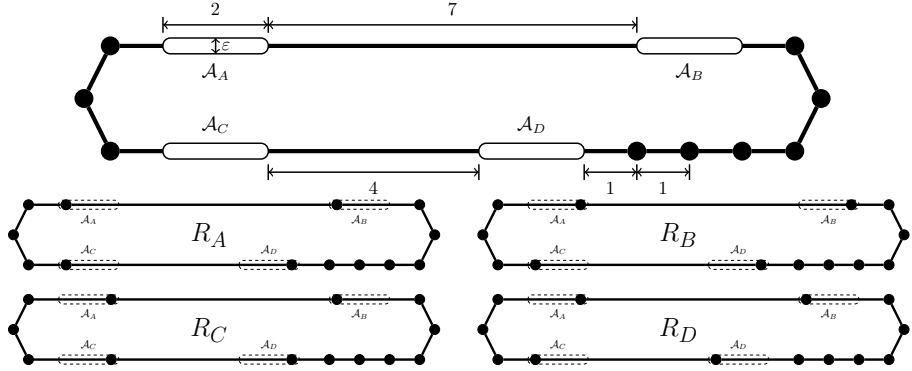}
	\caption{Instances $R_A$, $R_B$, $R_C$ and $R_D$ for the finite randomized family $(\mathcal{R},p)$ of instances}
	\label{v_mst_u}
\end{figure}

\begin{proof}
	We will prove the theorem by applying a variant of Yao's Principle (see \cite{borodin}) which allows to derive a lower bound for the performance of randomized algorithms from the best possible expected performance of a deterministic algorithm against a finite randomized family of instances.	
	A \emph{finite randomized family of instances} is a pair $(\mathcal{R},p)$ where $\mathcal{R}$ is a finite set of instances for V-MST-U and $p$ is a probability vector of length $\vert \mathcal{R} \vert$ which indicates for each instance $R \in \mathcal{R}$ the probability of its occurance. \\ 
	In \cite{verification}, Erlebach and Hoffmann define a graph $G$ with four non-trivial vertices $A$, $B$, $C$ and $D$. For each such vertex $v \in \{A,B,C,D\}$ they provide an instance $R_v$ where vertex positions are such that $\{v\}$ is the only optimal query set while after having queried any combination $S\subset \{A,B,C,D\} \setminus \{v\}$ of the other vertices, vertex positions can still turn out to be as in $R_u$ with $u\in \{A,B,C,D\} \setminus S$. For more details, see Figure \ref{v_mst_u}.\\

	\noindent Consider now the randomized family of instances $(\mathcal{R},p)$, where \\$\mathcal{R}=\{R_A, R_B, R_C, R_D\}$ and $p(R_v)=\mathbb{P}[R=R_v]=0.25$ for $v\in \{A,B,C,D\}$. Then no deterministic algorithm $ALG$ achieves a better expected competitive ratio 
	$\mathbb{E}_{R \sim_p \mathcal{R}}\left(\frac{ALG(R)}{OPT(R)}\right)$ than the algorithm $ALG_1$ which queries $A$, $B$, $C$, $D$ (or less if an MST can already be identified) in this order independently from the queries' results. This is due to the fact that querying $v$ only reveals whether or not we are facing instance $R_v$ but if not, it is indistinguishable which of the remaining instances it might be. 
	\noindent Then by Yao's Principle, no randomized algorithm has a better performance than 	
	\begin{alignat*}{2}
		&\min_{ALG \in \mathcal{A}} \mathbb{E}_{R \sim_p \mathcal{R}}\left(\frac{ALG(R)}{OPT(R)}\right) = \mathbb{E}_{R \sim_p \mathcal{R}}\left(\frac{ALG_1(R)}{OPT(R)}\right) &&= \\ &0.25\cdot\biggl(ALG_1(R_A)+ALG_1(R_B)+ALG_1(R_C)+ALG_1(R_D)\biggr) &&=\\&0.25\cdot(1+2+3+4)=2.5,
	\end{alignat*} where $\mathcal{A}$ denotes the class
	of all deterministic algorithms.
\end{proof}

\subsection{A randomized algorithm for V-MST-U on cactus-like graphs}

V-MST-U structurally differs from MST-U mainly due to the following two aspects: 
\begin{enumerate}[label=\alph*)]
	\item Querying a single end vertex of an edge already yields partial information about the edge's length. \label{aspecta}
	\item A vertex may be incident to several edges. Thus, knowing its precise position impacts on the possible lengths of all adjacent edges.\label{aspectb}
\end{enumerate}
While for MST-U it is sometimes possible to identify edges which have to lie in any feasible query set, it becomes significantly harder to tell whether a vertex has to be queried due to aspect \ref{aspecta}. However, we will make use of the weaker statement that for certain edges at least one out of two end vertices needs to be queried. \\
Similarly to the special case of cactus graphs for MST-U, we will now consider instances of V-MST-U where no two cycles share a non-trivial vertex and query costs are uniform. This makes it easier to deal with aspect \ref{aspectb}, as vertex queries may only impact on the possible lengths of at most two adjacent edges. We will refer to vertex-uncertainty graphs where cycles do not intersect in non-trivial vertices as \emph{cactus-like}.
Our algorithm considers cycles separately. For an instance $\mathcal{I}$ with a cycle $C$, our algorithm works as follows: First, it computes the associated edge instance and preprocesses the instance s.t. $T_L=T_U$ holds. Then it identifies an edge $f$ with largest upper limit $U_f$ in $C$. Usually, our algorithm deterministically queries the end vertices of $f$ and will then query the end vertices of edges in $C$ in order of decreasing upper limit in the associated edge instance until a longest edge in $C$ can be identified.   Only if the neighbor set is small (where ``small'' is yet to be defined) we will make use of randomization. In case that the neighbor set is large and the algorithm performs deterministically, we will say that the algorithm follows the \emph{deterministic procedure}. \\

\noindent \textbf{Remark:} To simplify notation, we will w.l.o.g. assume throughout this section that all uncertainty sets in associated edge instances have the form of intervals. However, the argumentation can be translated straightforwardly to the general case, e.g. by replacing $A_f$ \emph{contains} $A_e$ by $L_f \leq L_e \leq U_e \leq U_f$ where one out of two consecutive inequalities is strict.  

\subsubsection*{Preprocessing}

If $T_L \neq T_U$ there is exactly one edge in $T_L \setminus T_U$ because $C$ is a cycle. Our preprocessing consists in repeatedly computing $T_L$ and $T_U$ for the associated edge instance and querying the end vertices of the edge in $T_L \setminus T_U$ until $T_L = T_U$. However, the property $T_L = T_U$ does not necessarily remain true throughout the algorithm as vertex queries have impact on the uncertainty sets of all adjacent edges.

\subsubsection*{Deterministic procedure}
We will now argue why the deterministic procedure makes at most $2\cdot OPT(\mathcal{I}) + 2$ queries. 

\begin{lemma} \label{det_proc}
	Let $\mathcal{I}$ be an instance of V-MST-U on a cycle $C$. Let $ALG$ be an algorithm which computes the associated edge instance and in each subsequent step queries the end vertices of an edge with largest upper limit in the associated edge instance until a longest edge is found. Then $ALG$ makes at most $2\cdot OPT(\mathcal{I}) + 2$ queries when applied to $\mathcal{I}$. \\Moreover, if $\mathcal{I}$ is such that $T_L=T_U$, $f_1 \in C-T_L$ is the first edge with largest upper limit considered by $ALG$ and $Q^*$ is an optimal query set with $f_1 \cap Q^* \neq \emptyset$ then $ALG$ makes at most $2.5\cdot OPT(\mathcal{I})$ queries when applied to $\mathcal{I}$.
\end{lemma}

\begin{proof}
	We will denote by $\{u_i,v_i\}$ the set of vertices that $ALG$ queries during the $i$'th step. (Note that $u_i=v_i$ is possible if the edge considered in the $i$'th step has one trivial end vertex.) Moreover, we will denote by $V_i$ the set of vertices queried during the first $i$ steps such that $\vert V_T \vert =ALG(\mathcal{I})$ where $T$ is the last iteration. Let $Q^*$ be an optimal query set and denote by $A^i_e$, $U^i_e$ and $L^i_e$ the uncertainty set, upper limit and lower limit of an edge $e$ in iteration $i$. 
	We will argue that for each iteration $i$ it either holds that at least one vertex in  $\{u_i,v_i\}$ lies in $Q^*$ or that for each $j<i$ at least one vertex in  $\{u_j,v_j\}$ lies in $Q^*$. This trivially holds for the first iteration. Assume now that it is known to be true for iterations $1,...,i-1$ and consider the $i$-th iteration. Note that $u_i$ and $v_i$ are non-trivial end vertices of an edge $f$ which has largest upper limit $U_f^i$ in the uncertainty graph considered in iteration $i$. First assume that there exists an edge $e\in C$ with $A_e^i \subset A_f^i$. Then by Lemma \ref{lem2_v} we need to know the position of at least one of $f$'s end vertices in order to be able to identify a minimum spanning tree. Now assume that none of the other uncertainty sets is a subset of $A_f^i$. By induction hypothesis we know that there is at most one $j>i$ for which we have not yet argued that $\{u_j,v_j\}$ intersects with $Q^*$. Let $h$ be the edge with largest upper limit in iteration $j$ such that $u_j$ and $v_j$ are end vertices of $h$. Then $w_h$ can neither lie in $A_f^i$ else $A_h^i=\{w_h\} \subset A_f^i$ nor can we have that $U_h ^i=w_h>U_f^i$ as $f$ has largest upper limit in the $i$-th iteration. Hence we know that $w_h <w_f$ and thus $h$ lies in any MST. Let $g$ be a longest edge in $C$. By the choice of $h$ we have that $U_h^j \geq U_g^j \geq w_g$ and thus without querying neither $u_j$ nor $v_j$ (which are the non-trivial end vertices of $h$ in iteration $j$) we will not be able to verify an MST which does not contain $g$. \\
	From this the first part of the lemma follows: Let $j$ be the last step for which we cannot guarantee that $v_j$ or $u_j$ lies in any feasible query set. Then we know that half of $V_{j-1}$ lies in $Q^*$ and that for each step $i>j$ we know that $v_i$ or $u_i$ lies in $Q^*$. Hence, $OPT(\mathcal{I})= \vert Q^* \vert \geq \frac{\vert V_{j-1}\vert}{2} + \frac{\vert V_{T} \vert - \vert V_{j}\vert}{2} = \frac{\vert V_{T} \vert - \vert \{u_j,v_j\}\vert}{2} \geq \frac{ALG(\mathcal{I}) - 2}{2}$. \\
	Note that if $j \neq 1$ then $u_j=v_j$: By the choice of $j$, no uncertainty set is contained in $A_e^j$ where $e$ is the edge with largest upper limit in iteration $j$ and non-trivial end vertices $u_j$ and $v_j$. Thus $w_{f_1} \leq L_e$ which cannot be the case if both end vertices of $e$ remain unqueried until iteration $j$ due to our preprocessing. Hence, the first iteration is the only iteration where we might query two vertices none of which is in $Q^*$ and thus if $Q^* \cap f_1 \neq \emptyset$ then $ALG$ makes at most $2\cdot OPT(\mathcal{I}) + 1$ queries, i.e. it achieves competitive ratio $2.5.$ unless $OPT(\mathcal{I})=1$. However, if $OPT(\mathcal{I})=1$ and $Q^*$ intersects $f_1$ then $Q^*$ is a subset of $f_1$ and thus the deterministic procedure finishes after querying only the end vertices of $f_1$. This proves the second part of the claim.

\end{proof}

\subsubsection*{Randomization}

The above lemma guarantees that the deterministic procedure achieves a competitive ratio of $2+\frac{2}{OPT(\mathcal{I})}$ which is at most $2.5$ unless $OPT(\mathcal{I})\leq 3$. We will first handle the exception where $OPT(\mathcal{I})=3$ by making a slight adaption: The algorithm starts with the first three iterations of the deterministic procedure. In the fourth iteration it queries a vertex in $\{u_4,v_4\}$ with probability $0.5$ (or probability $1$ if $u_4=v_4$), then queries the other vertex if necessary and finally proceeds with the deterministic procedure if still no longest edge can be identified. See Algorithm \ref{alg_rand3} for a precise description of the algorithm \mysc{Rand3} that is  applied to the uncertainty graph that results from querying $f_1$'s end vertices.
\begin{lemma} \label{rand3}
	Let $\mathcal{I}$ be a preprocessed instance of V-MST-U on a cycle $C$. Let $ALG$ be an algorithm which computes the associated edge instance,  queries the end vertices of an edge $f_1$ with largest upper limit in the associated edge instance and applies \mysc{Rand3} to the resulting uncertainty graph. Then $ALG$ achieves competitive ratio $2.5$ when applied to $\mathcal{I}$ unless $OPT(\mathcal{I}) \leq 2$ and $f_1 \cap Q^* = \emptyset$ for all optimal query sets $Q^*$. 
\end{lemma}

\begin{proof}
	Again we denote by $\{u_i,v_i\}$ the set of vertices that $ALG$ queries during the $i$'th step (i.e. $\{u_1,v_1\}=f_1$). The algorithm $ALG$ makes at most as many queries as the deterministic procedure. Thus if $Q^* \cap f_1 \neq \emptyset$ or  $OPT(\mathcal{I}) \geq 4$ then the claim follows from Lemma \ref{det_proc}. Assume now that $Q^* \cap f_1 = \emptyset$ and $OPT(\mathcal{I})=3$. If the algorithm finishes after the first three iterations then it has made at most $6$ queries and yields a competitive ratio of at most $2$. If the algorithm has not finished after the first three iterations, then we know from the proof of Lemma \ref{det_proc} that two of the three sets $\{u_1,v_1\}$ $\{u_2,v_2\}$ and $\{u_3,v_3\}$ intersect with $Q^*$ such that $\{u_4,v_4\}$ contains the last vertex in $Q^*$. Hence, we expect to make $0.5 \cdot (7+8)$ queries which yields a competitive ratio of at most $2.5$.
\end{proof}

\IncMargin{1em}
\begin{algorithm}[H]
	\SetKwData{Left}{left}\SetKwData{This}{this}\SetKwData{Up}{up}
	\SetKwFunction{Union}{Union}\SetKwFunction{FindCompress}{FindCompress}
	\SetKwInOut{Input}{input}\SetKwInOut{Output}{output}
	\Input{An instance $\mathcal{I}$ of V-MST-U with uniform query costs with a cycle $C=(V,E)$ and uncertainty sets $A_v$ for $v \in V$}
	\Output{A feasible query set $Q$}
	\BlankLine
	Compute the associated edge instance and $T_L$\;
	Initialize $Q=\emptyset$\;
	\For{$i\leftarrow 1$ \KwTo $2$}{
		
		\If{no edge in $C$ is always maximal}{Let $g$ be an edge with largest upper limit in $C$\; Query the end vertices of $g$ and add them to $Q$}}
	\If{no edge in $C$ is always maximal}{Let $g$ be an edge with largest upper limit in $C$ and let $u,v$ be the non-trivial end vertices of $g$\; Pick $b\in [0,1]$ uniformly at random\;
		\If{$b\leq 0.5$}{Query $u$ and add it to $Q$\;
			\If{no edge in $C$ is always maximal}{Query $v$ and add it to $Q$} }
		\Else{Query $v$ and add it to $Q$\;  \If{no edge in $C$ is always maximal}{Query $u$ and add it to $Q$}} 
	}
	\While{no edge in $C$ is always maximal}{Let $g$ be an edge with largest upper limit in $C$\; Query the end vertices of $g$ and add them to $Q$}
	\caption{The subroutine \mysc{Rand3} of \mysc{V-Random}$_C$}\label{alg_rand3}
\end{algorithm}\DecMargin{1em}

\vspace{.5cm}\noindent Following from Lemma \ref{rand3}, we only need to deal with instances where $OPT(\mathcal{I}) \leq 2$ and $f_1 \cap Q^* = \emptyset$. Thus we first check whether the (original) uncertainty graph contains candidates for a feasible query set $W$ of size at most $2$ that does not contain an end vertex of $f_1$. We do so by computing the size $a$ of a smallest vertex cover of the non-trivial edges in $X(f_1)$ which does not intersect $f_1$ and consists of non-trivial vertices only. Note that if there is a feasible query set of size at most $2$ which does not intersect the edge $f_1$ then $a \leq 2$ by Lemma \ref{lem2_v}.\\
If $a=1$ we apply a randomized procedure \mysc{Rand1} as described below. If $a=2$, we query the end vertices of $f_1$ and depending on the outcome either decide to stick to the strategy \mysc{Rand3} or apply a randomized strategy as described below. 
\IncMargin{1em}
\begin{algorithm}[ht]
	\SetKwData{Left}{left}\SetKwData{This}{this}\SetKwData{Up}{up}
	\SetKwFunction{Union}{Union}\SetKwFunction{FindCompress}{FindCompress}
	\SetKwInOut{Input}{input}\SetKwInOut{Output}{output}
	\Input{An instance $\mathcal{I}$ of V-MST-U with uniform query costs with a cycle $C=(V,E)$ and uncertainty sets $A_v$, $v \in V$}
	\Output{A feasible query set $Q$}
	\BlankLine
	Compute the associated edge instance\;
	Initialize $Q=\emptyset$\;
	Let $f_1$ be an edge with largest upper limit and let $X(f_1)$ be the set of edges $g \in T_L \cap C$ with $U_g > L_{f_1}$\;
	\If{no edge in $C$ is always maximal}{
		
		index the vertices $v_1,...,v_{k}$ in $\bigcap X(f_1) \cup f_1$ arbitrarily\;
		index the permutations $\sigma_1,...,\sigma_{k!}$ in $\mathcal{S}_k$ arbitrarily\;
		draw $b$ uniformly at random from $[0,1]$ and pick permutation  $\sigma=\sigma_i$ if $b\in \left[\left.\frac{i-1}{k!},\frac{i}{k!}\right.\right)$\;
		$j:=1$\;
		\While{no edge in $C$ is always maximal and $j<k$}{Query $v_{\sigma_i(j)}$ and add it to $Q$ \; $j:=j+1$\;} 
		\If{no edge in $C$ is always maximal}{Query the remaining vertices in $X(f_1)$}}
	
	\caption{The subroutine \mysc{Rand1} of \mysc{V-Random}$_C$}\label{alg_rand1}
\end{algorithm}\DecMargin{1em}

\IncMargin{1em}
\begin{algorithm}[ht]
	\SetKwData{Left}{left}\SetKwData{This}{this}\SetKwData{Up}{up}
	\SetKwFunction{Union}{Union}\SetKwFunction{FindCompress}{FindCompress}
	\SetKwInOut{Input}{input}\SetKwInOut{Output}{output}
	\Input{An instance $\mathcal{I}$ of V-MST-U with uniform query costs with a cycle $C=(V,E)$ and uncertainty sets $A_v$, $v \in V$}
	\Output{A feasible query set $Q$}
	\BlankLine
	Compute the associated edge instance and $T_L$\;
	Initialize $Q=\emptyset$\;
	Let $f_1$ be the edge in $R := E \setminus T_L$ and let $X(f_1)$ be the set of edges $g \in T_L \cap C$ with $U_g > L_{f_1}$\;
	Query $f_1$ and add its end vertices to $Q$\;
	Let $g$ be an edge with largest upper limit in $C$\;
	\If{no edge in $C$ is always maximal}{
		\If{$w_{f_1}\leq L_g$}{$Q=Q \cup$ \mysc{Rand3}$(C_i, \{A_v\vert v \in V(C_i)\})$}
		\Else{
			Compute $\mathcal{W}=\{W\subset V(C)-f_1 \vert$ W is a smallest vertex cover of $X(f_1)$ which consists of non-trivial vertices only$\}$ and index the elements $W_1,...,W_{\vert \mathcal{W} \vert}$ arbitrarily\;
			draw $b$ uniformly at random from $[0,1]$ and pick a vertex cover $W=W_i$ if $b\in \left[\left.\frac{i-1}{\vert \mathcal{W}\vert},\frac{i}{\vert \mathcal{W}\vert}\right.\right)$\;
			query a vertex $v$ in $W \cap g$ and add it to $Q$\;
			\If{no edge in $C$ is always maximal}{
				query a vertex in $W-v$ and add it to $Q$}
			\While{no edge in $C$ is always maximal}{Let $h$ be an edge with largest upper limit in $C$\; query the end vertices of $h$ and add them to $Q$} }
	}
	\caption{The subroutine \mysc{Rand2} of \mysc{V-Random}$_C$}\label{alg_rand2}
\end{algorithm}\DecMargin{1em}

\begin{figure}[h!]
	\centering
		\scalebox{1}{ 
\begin{tikzpicture} 
	\node (1) at (1,0) [circle,draw] {};
	\node (a) at (1.5,-0.5) [] {(a)};
	\node (b) at (4,-0.5) [] {(b)};
	\node (c) at (6.5,-0.5) [] {(c)};
	\node (2) at (2,0) [circle,draw, minimum size = 9pt] {};
	\node (20) at (2,0) [] {\scriptsize $w_2$};
	\node (3) at (3.5,0) [circle,draw, minimum size = 9pt] {};
	\node (30) at (3.5,0) [] {\scriptsize $w_3$};
	\node (4) at (4.5,0) [circle,draw, minimum size = 9pt] {};
	\node (40) at (4.5,0) [] {\scriptsize $w_2$};
	\node (5) at (6,0) [circle,draw, minimum size = 9pt] {};
	\node (50) at (6,0) [] {\scriptsize $w_4$};
	\node (6) at (7,0) [circle,draw, minimum size = 9pt] {};
	\node (60) at (7,0) [] {\scriptsize $w_2$};
	\node (9) at (1,1) [circle,draw, minimum size = 9pt] {};
	\node (10) at (2,1) [circle,draw, minimum size = 9pt] {};
	\node (100) at (2,1) [] {\scriptsize $w_1$};
	\node (11) at (3.5,1) [circle,draw, minimum size = 9pt] {};
	\node (12) at (4.5,1) [circle,draw, minimum size = 9pt] {};
	\node (120) at (4.5,1) [] {\scriptsize $w_1$};
	\node (13) at (6,1) [circle,draw, minimum size = 9pt] {};
	\node (130) at (6,1) [] {\scriptsize $w_3$};
	\node (14) at (7,1) [circle,draw, minimum size = 9pt] {};
	\node (140) at (7,1) [] {\scriptsize $w_1$};
	\node (17) at (.25,.5) [circle,draw, minimum size = 9pt] {};
	\node (170) at (.25,.5) [] {\scriptsize $w_3$};
	\draw[-,dashed] (1) to (2);
	\draw[-, line width = 2 pt] (3) to (4);
	\draw[-,dashed] (5) to (6);

	\draw[-,dashed] (11) to (12);
	\draw[-, line width = 2 pt] (1) to (17);
	\draw[-, line width = 2 pt] (9) to (17);
	\draw[-,dashed] (13) to (14);

	\draw[-] (2) to (10);
	\draw[-,dashed] (9) to (10);
	\draw[-, dashed] (3) to (11);
	\draw[-] (4) to (12);
	\draw[-, line width = 2 pt] (5) to (13);
	\draw[-] (6) to (14);

	\path  (2) -- (10) node[draw=none, midway, right=4pt]{$f_1$};
	\path  (4) -- (12) node[draw=none, midway, right=4pt]{$f_1$};
	\path  (6) -- (14) node[draw=none, midway, right=4pt]{$f_1$};
\end{tikzpicture}}\caption{Sketches of cycles with $a=1$ where edges in $X(f_1)$ are bold and dashed lines indicate parts of the cycle that are not in $X(f_1) \cup f_1$. In (a) and (b) \mysc{Rand1} picks a vertex in $\{w_1,w_2,w_3\}$ with probability $\frac{1}{3}$ each. In (c) \mysc{Rand1} picks a vertex in $\{w_1,w_2,w_3,w_4\}$ with probability $\frac{1}{4}$ each.}\label{a_1} 	
\end{figure}
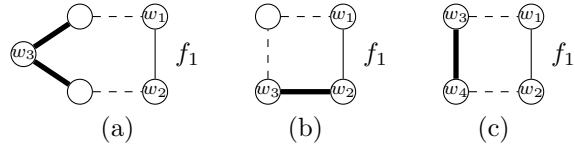

\noindent \textbf{Case }$\mathbf{a = 1:}$ If $a=1$ then $X(f_1)$ either consists of a single edge $e$ or of two edges $g$ and $h$ which are incident to each other but not to $f_1$ (see Figure \ref{a_1}). The algorithm picks an order in which to query the three (or four) vertices in $(g \cap h) \cup f_1$ or $e  \cup f_1$ uniformly at random and queries the vertices in the respective order up to the point where a longest edge in $C$ can be identified. If necessary, it queries the remaining vertices in $X(f_1)$. See Algorithm \ref{alg_rand1} for a precise description.\\
\textbf{Case }$\mathbf{a = 2:}$ The algorithm starts by querying the end vertices of $f_1$. Let $g$ be an edge in the resulting uncertainty graph with largest upper limit $U_g$. If now $w_{f_1} \leq L_g$ then we continue as in \mysc{Rand3}. If $w_{f_1} \in A_g$ then the algorithm picks a vertex cover $W$ of  $X(f_1)$ which contains $\vert W \vert = 2$ non-trivial vertices such that $W \cap f_1 = \emptyset$ uniformly at random. The algorithm queries the vertices in $W$ but starts with a vertex in $W \cap g$. After that it queries the end vertices of edges in order of decreasing upper limit. For a precise description, see Algorithm \ref{alg_rand2}.

\begin{lemma} \label{rand2}
	Let $\mathcal{I}$ be a preprocessed instance of V-MST-U on a cycle $C$. Let $f_1$ be an edge with largest upper limit in the associated edge instance and denote by $a$ the size of a smallest vertex cover of $X(f_1)$ which consists of non-trivial vertices only. If $a=2$ then Algorithm \ref{alg_rand2} achieves competitive ratio $2.5$ when applied to $\mathcal{I}$.
\end{lemma}

\begin{proof}
	As $\vert f_1 \cup V(X(f_1)) \vert \leq 8$ (see Figure \ref{a_2}) we only need to discuss instances with $OPT(\mathcal{I})=\vert Q^* \vert < 4$ where $Q^*$ is an optimal query set. If $OPT(\mathcal{I})=1$ then $a=2$ implies that $Q^* \subset f_1$ by Lemma \ref{lem2_v} and the vertex in $Q^*$ is queried within the first two queries. Say $OPT(\mathcal{I})=2$. Let $g$ be an edge with largest upper limit in the associated edge instance after having queried the end vertices of $f_1$. We will distinguis the cases where $Q^*$ either intersects $f_1$ or not. First assume that $Q^* \cap f_1 \neq \emptyset$. If $Q^* = f_1$ then the claim trivially holds. We thus assume that $\vert Q^* \cap f_1 \vert = 1$. If $w_{f_1} \leq L_g$ we proceed as in \mysc{Rand3} and the claim follows from Lemma \ref{rand3}. Thus, consider the case where $w_{f_1} \in A_g$. \mysc{Rand2} proceeds by querying a vertex cover of $X(f_1)$. We will now argue that if still no longest edge is found and $h$ is an edge which subsequently has largest upper limit in $C$ then the remaining vertex $v \in Q^*$ is the (unique) non-trivial end vertex of $h$. If $Q^*$ verifies an MST which contains $h$ then $Q^*$ contains the end vertex of $h$ otherwise $h$ will remain a candidate for a strictly longest edge. Note that $h$ is not incident to $f_1$ otherwise $h$ would be trivial after querying $f_1$ and a vertex cover of $X(f_1)$. Thus after querying $f_1$ we have that $w_{f_1} \in A_h$ due to the preprocessing which proves that $v$ needs to be an end vertex of $h$ if $Q^*$ verifies that $h$ is a longest edge in $C$. Hence, if $OPT(\mathcal{I}) = 2$ and $Q^* \cap f_1 \neq \emptyset$ we can verify an MST after querying the end vertices of $f_1$, a vertex cover of size $2$ and an end vertex of $h$ which proves a competitive ratio of at most $2.5$.\\
	\begin{figure}
		\centering 
		\begin{tikzpicture} 
		\node (b) at (2,-2) [] {\text{(b)}};
		\node (1) at (.75,-.25) [circle,draw, minimum size = 9pt] {};
		\node (2) at (1.5,0.25) [circle,draw, minimum size = 9pt] {};

		\node (5) at (2.5,.25) [circle,draw, minimum size = 9pt] {};
		\node (6) at (3.25,-.25) [circle,draw, minimum size = 9pt] {};
		\node (7) at (3.25,-1) [circle,draw, minimum size = 9pt] {};
		\node (70) at (3.25,-1) [inner sep=0pt] {\scriptsize $w_2$};
		\node (8) at (2.5,-1.5) [circle,draw, minimum size = 9pt] {};
		\node (9) at (1.5,-1.5) [circle,draw, minimum size = 9pt] {};
		\node (10) at (.75,-1) [circle,draw, minimum size = 9pt] {};
		\node (100) at (.75,-1) [inner sep=0pt] {\scriptsize $w_1$};
		\draw[-, dashed] (1) to (2);
		
		\path[draw, line width= .8 pt] (2) -- (5) node[draw=none, midway, above =4pt]{$f_1$};
		\draw[-, dashed] (5) to (6);
		\draw[-, line width= 2 pt] (6) to (7);
		\draw[-, line width= 2 pt] (7) to (8);
		\draw[-, dashed] (8) to (9);
		\draw[-, line width= 2 pt] (9) to (10);
		\draw[-, line width= 2 pt] (10) to (1);
		\draw[-,dashed ]  (8) to (9);
		
		\node (a) at (-2,-2) [] {\text{(a)}};
	
		\node (2a) at (-2.5,0.25) [circle,draw, minimum size = 9pt] {};
		
		\node (5a) at (-1.5,.25) [circle,draw, minimum size = 9pt] {};
		
		\node (7a) at (-.75,-.6251) [circle,draw, minimum size = 9pt] {};
		\node (70a) at (-.75,-.6251) [inner sep=0pt] {\scriptsize $w'_1$};
		\node (8a) at (-1.5,-1.5) [circle,draw, minimum size = 9pt] {};
		\node (80a) at (-1.5,-1.5) [inner sep=0pt] {\scriptsize $w'_2$};
		\node (9a) at (-2.5,-1.5) [circle,draw, minimum size = 9pt] {};
		\node (90a) at (-2.5,-1.5) [inner sep=0pt] {\scriptsize $w_2$};
		\node (10a) at (-3.25,-.6251) [circle,draw, minimum size = 9pt] {};
		\node (100a) at (-3.25,-.6251) [inner sep=0pt] {\scriptsize $w_1$};
		\draw[-, dashed] (10a) to (2a);
		
		\path[draw, line width= .8 pt] (2a) -- (5a) node[draw=none, midway, above =4pt]{$f_1$};
		\draw[-, dashed] (5a) to (7a);
	
		\draw[-, line width= 2 pt] (7a) to (8a);
		\draw[-, dashed] (8a) to (9a);
		\draw[-, line width= 2 pt] (9a) to (10a);
		
		\draw[-,dashed ]  (8a) to (9a);
\end{tikzpicture}\caption{Sketches of cycles with $a=2$ where edges in $X(f_1)$ are bold and dashed lines indicate parts of the cycle that are not in $X(f_1) \cup f_1$. In \text{(a)} \mysc{Rand2} picks a cover $\{w_i,w'_j\}$, $i,j=1,2$ with probability $\frac{1}{4}$ each. In \text{(b)} \mysc{Rand2} picks $\{w_1,w_2\}$ deterministically.}\label{a_2} 	
	\end{figure}
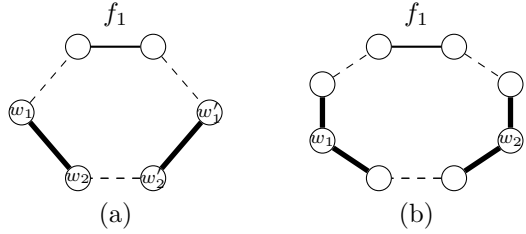Assume now that $Q^* \cap f_1 = \emptyset$. By Lemma \ref{lem2_v}, this implies that $Q^*$ is a vertex cover of $X(f_1)$ (which verifies that $f_1$ is a longest edge in $C$). Say \mysc{Rand2} picks a vertex cover which is not a feasible query set. During any of the following steps, let $g$ be an edge which currently has largest weight in $C$. Then $A_g$ contains $w_{f_1}$ and thus $Q^*$ contains the non-trivial vertex of $g$ else $g$ cannot be shown to have at most the weight of $f_1$. Note that we can have at most four different smallest vertex covers of $X(f_1)$ of which do not intersect $f_1$ and at most two which are pairwise disjoint from each other (see Figure \ref{a_2}). Thus with probabilty at most $\frac{1}{4}$ we pick $Q^*$ right away, with probability $\frac{1}{4}$ we pick a vertex cover that is disjoint from $Q^*$ and need two additional queries while with probability $\frac{1}{2}$, we pick a vertex cover that intersects with $Q^*$ in a single vertex after which we need to make one more query. This leads to a competitive ratio of $\frac{0.25\cdot 4+ 0.25 \cdot 6+0.5\cdot 5}{2}= 2.5$. \\
	To complete the proof, we need to discuss instances with $OPT(\mathcal{I})=3$. Note that the claim trivially holds if we have that $\vert f_1 \cup V(X(f_1)) \vert >7$ which is the case unless $X(f_1)$ consists of two paths which are neither incident to each other nor to $f_1$ (as in Figure \ref{a_2} \text{(b)} ). Assume that the latter is the case and that we have queried the end vertices of $f_1$ and the vertices in the vertex cover $W$ of $X(f_1)$. Thus, all remaining vertices in $X(f_1)$ have at most one non-trivial end vertex. As $V(C)-W-f_1$ contains no such vertex cover of $X(f_1)$ of size $3$, Lemma \ref{lem2_v} guarantees that we have already queried one of the vertices in $Q^*$ before we consider the edge $h$. Analagously to the proof of Lemma \ref{det_proc} for the deterministic procedure, we can argue that at least two out of the next three queries contain vertices in $Q^*$ and thus we make at most seven queries in total which proves the claim for this case.  
\end{proof}
\subsubsection{The algorithm \mysc{V-Random}$_C$}

A precise description of the algorithm \mysc{V-Random}$_C$ is displayed in Algorithm \ref{vrandomc}

\IncMargin{1em}
\begin{algorithm}[ht]
	\SetKwData{Left}{left}\SetKwData{This}{this}\SetKwData{Up}{up}
	\SetKwFunction{Union}{Union}\SetKwFunction{FindCompress}{FindCompress}
	\SetKwInOut{Input}{input}\SetKwInOut{Output}{output}
	\Input{An instance $\mathcal{I}$ of V-MST-U with uniform query costs, a graph $G=(V,E)$ and uncertainty sets $A_v$, $v \in V$ such that no two cycles share a non-trivial vertex}
	\Output{A feasible query set $Q$}
	\BlankLine
	Compute the associated edge instance\;
	Initialize $Q=\emptyset$\;
	Preprocess the instance such that $T_L=T_U$ \;
	Index the edges $f_1, . . . , f_{m-n+1}$ in $R := E \setminus T_L$ arbitrarily \;
	
	\For{$i\leftarrow 1$ \KwTo $m-n+1$}{
		Let $C_i$ be the unique cycle in $T_L + f_i$\;
		Let $X(f_i)$ be the set of edges $g \in T_L \cap C_i$ with $U_g > L_{f_i}$\;
		Compute the size $a$ of a smallest vertex cover of $X(f_1)$ which does not intersect $f_1$ and consists of non-trivial vertices only\;
		\If{no edge in $C_i$ is always maximal}{
			\If{$a=1$}{$Q=Q \cup$ \mysc{Rand1}$(C_i, \{A_v\vert v \in V(C_i)\})$}
			\Else{Query the end vertices of $f_i$ and add them to $Q$\;
				Let $g$ be an edge with largest upper limit in $C_i$\;
				\If{$a \geq 3$ or $w_{f_i} \leq L_g$}{$Q=Q \cup$ \mysc{Rand3}$(C_i, \{A_v\vert v \in V(C_i)\})$}
				\Else{$Q=Q \cup$ \mysc{Rand2}$(C_i, \{A_v\vert v \in V(C_i)\})$}}
	}}
	\caption{The algorithm \mysc{V-Random}$_C$ for V-MST-U with uniform query costs in cactus-like graphs}\label{vrandomc}
\end{algorithm}\DecMargin{1em}

\begin{theorem}
	For instances of V-MST-U with uniform query cost where no two cycles intersect in a non-trivial vertex the algorithm \mysc{V-Random}$_C$ achieves a competitive ratio of 2.5 and this is best possible.
\end{theorem}

\begin{proof}
	
	Let $\mathcal{I}$ be an instance with uncertainty graph $G$ to which we apply \mysc{V-Random}$_C$. For a cycle $C$ in $G$ we denote by $Q_C^*$ an optimal query set for an instance of V-MST-U with graph $C$ and uncertainty sets and weights as in $\mathcal{I}$. If we prove for each such cycle $C$ that \mysc{V-Random}$_C$ achieves a competitive ratio of $2.5$ when applied to $C$ the claim follows because $Q^*=\dot\bigcup_C Q_C^* $ is an optimal query set for $\mathcal{I}$. Moreover, Lemma \ref{preproc_v} guarantees that each vertex pair queried during the preprocessing intersects with $Q^*$.
	Thus we assume that $\mathcal{I}$ is a preprocessed instance of V-MST-U with uncertainty graph $C$ where $C$ is a cylce. Let $Q^*$ be an optimal query set for $\mathcal{I}$. Now we prove that \mysc{V-Random}$_C$ queries at most $2.5\cdot OPT(\mathcal{I})$ vertices when applied to $\mathcal{I}$. \\
	First assume that $a \geq 3$, i.e. the algorithm queries the end vertices of $f_1$ and applies \mysc{Rand3} to the resulting uncertainty graph. From  $a \geq 3$ it follows by Lemma \ref{lem2_v} that $Q^* \cap f_1 \neq \emptyset$ or $OPT(\mathcal{I}) \geq a \geq 4$ and thus the claim follows from Lemma \ref{rand3}.\\ 
	if we assume that $a=2$ then the claim follows from Lemma \ref{rand2}. 
	Finally, assume that $a=1$. Note that the maximum number of vertices in $f_1 \cup V(X(f_1))$ is $5$. Thus we make at most five queries and only need to consider the case where $OPT(\mathcal{I})=1$. The neighbor set $X(f_1)$ consists either of a single edge $e$ or of two intersecting edges $g$ and $h$. In the first case, each non-trivial vertex in $e \cup f_1$ could be a feasible query set of size one and we find the vertex in $Q^*$ within at most $\frac{1}{4}(1+2+3+4)=2.5$ queries in expectation. In the second case each non-trivial vertex in $(g \cap h) \cup f_1$ could be a feasible query set of size one and we find the vertex in $Q^*$ within $\frac{1}{3}(1+2+3)=2$ queries in expectation. Note that if $a=0$ then $f_i$ has no neighbors and is thus always maximal right away. The optimality follows from Theorem \ref{vertexbound_d} and the fact that all uncertainty graphs used throughout its proof are cycles.
	
\end{proof}

\section{Minimum Spanning Stars}

Several online algorithms exist for MST-U and are shown to deliver a solution the cost of which is bounded by a constant multiple of the cost of an optimal solution. But what happens if we ask to find a minimum spanning tree of a specified type at minimum query cost? For many types of special spanning trees, it is already NP-hard to find such a tree in a graph, e.g. Hamiltonian paths or spiders (see Gargano et al. \cite{gargano}). In the following we will thus consider spanning stars which are generally easy to find.

\begin{definition}
	The complete bipartite graph $K_{1,k}$ is called a \emph{star} and is denoted by $S_k$. We will refer to the vertex with degree $k$ as centre of $S_k$. If a graph $G$ with $n$ vertices contains $S_{n-1}$ as a subgraph, then $S_{n-1}$ is said to be a spanning star in $G$.  
\end{definition}

\noindent We define the \emph{Minimum Spanning Star Problem under Explorable Uncertainty} (MSS-U) analogously to MST-U, i.e. we want to identify a minimum spanning star by making queries of as little weight as possible. However, it turns out that no algorithm for MSS-U can achieve constant competitive ratio. 

\begin{theorem}
	There exists no algorithm for MSS-U which achieves constant competitive ratio. 
\end{theorem}

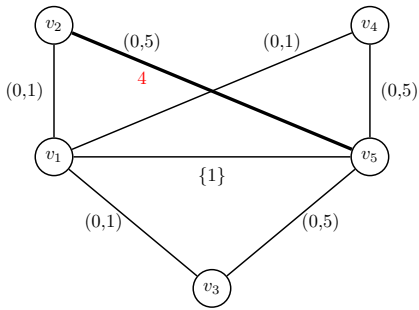
\begin{figure}[h!]
	\centering
	\scalebox{.7}{
		\begin{tikzpicture}[thick]
			\node (1) at (0,0) [circle,draw] {$v_1$};
			\node (5) at (6,0) [circle,draw] {$v_5$};
			\node (4) at (6,2.5) [circle,draw] {$v_4$};
			\node (3) at (3,-2.5) [circle,draw] {$v_3$};
			\node (2) at (0,2.5) [circle,draw] {$v_2$};
			\draw[-] (1) to (2);
			\draw[-] (1) to (3);
			\draw[-] (1) to (4);
			\draw[-] (1) to (5);
			\draw[-, line width= 1.8 pt] (5) to (2);
			\draw[-] (5) to (3);
			\draw[-] (5) to (4);
			\path  (1) -- (5) node[draw=none, midway, below=2pt]{$\{1\}$};
			\path  (1) -- (4) node[draw=none, near end, above=2pt]{(0,1)};
			\path  (1) -- (3) node[draw=none, midway, left=2pt]{(0,1)};
			\path  (1) -- (2) node[draw=none, midway, left=2pt]{(0,1)};
			\path  (5) -- (4) node[draw=none, midway, right=2pt]{(0,5)};
			\path  (5) -- (3) node[draw=none, midway, right=2pt]{(0,5)};
			\path  (5) -- (2) node[draw=none, near end, above=2pt]{(0,5)};
			\path  (5) -- (2) node[draw=none, near end, below=2pt, color =red]{4};
	\end{tikzpicture}}
	\caption{The weight of $\{v_2,v_5\}$ is $4$. All missing edge weights equal $0.5$. An optimal solution only needs to query $\{v_2,v_5\}$ which has larger weight than the star with center $v_1$ can have. A deterministic algorithm can not distinguish between the edges $\{v_2,v_5\}, \{v_3,v_5\}$ and $\{v_4,v_5\}$ and might have to query all three of them.}\label{star}
\end{figure}

\begin{proof}
	We prove the theorem by defining for each $n \in \mathbb{N}$ a finite family $(\mathcal{I}_n^k)_k$ of instances with uniform query cost on a graph $G$ with $n$ vertices such that every algorithm needs to make at least $(n-2)$ times as many queries as necessary on at least one instance in $(\mathcal{I}_n^k)_k$. For $n=5$ the construction is illustrated in Figure \ref{star}.\\
	Consider a graph $G=(V,E)$ with $V=\{v_1,...,v_n\}$ and $E=\{\{v_1,v_i\} \vert i=2,...,n\} \cup \{\{v_n,v_i\} \vert i=1,...,n-1\}$. Then $G$ has precisely two spanning stars, one with centre $v_1$ and the other one with centre $v_n$. (All other vertices have degree $2$.) Let $q_e=1$ for all $e \in E$. We define the uncertainty sets as follows:
	\begin{itemize}
		\item $A_{\{v_1,v_n\}}=\{1\}$,
		\item $A_{\{v_1,v_i\}}= (0,1)$ for $i=2,...,n-1$ and 
		\item $A_{\{n,v_i\}}=(0,n)$ for $i=2,...,n-1$.
	\end{itemize}
	
	\noindent For $j\in\{2,...,n-1\}$, $\mathcal{I}_n^j$ is defined such that $w_{\{v_j,v_n\}}=n-1$ and $w_{e} = 0.5$ for all $e \in E \setminus \{\{v_j,v_n\}, \{v_1,v_n\}\}$. It is then sufficient to query only the edge $\{v_j,v_n\}$ to know that the spanning star with center $v_n$ has larger weight. Conversely, without querying $\{v_j,v_n\}$, it is impossible to tell which spanning star has minimum weight. Note that for an algorithm all non-trivial edges adjacent to $v_n$ are undistinguishable. It has thus to decide on an order in which these edges are queried. If $j$ is such that $\{v_j,v_n\}$ is the last edge adjacent to $v_n$ to be queried by an algorithm then the algorithm competitive ratio at least $n-2$ when applied to instance $\mathcal{I}_n^j$. 
\end{proof}

\section{Conclusion}
In this paper we considered the Minimum Spanning Tree Problem Under Explorable Uncertainty (MST-U) and the related Minimum Spanning Tree Problem Under Explorable Vertex Uncertainty (V-MST-U) for specified instance types where cycles can be considered independently from each other. We provided a randomized algorithm for MST-U on cactus graphs and proved that it achieves a competitive ratio of $1.5$ which is best possible. For V-MST-U instances where cycles do not share non-trivial vertices and query costs are uniform, we provided the algorithm \mysc{V-Random}$_C$ which achieves a competitive ratio of $2.5$. We showed that $2.5$ is a lower bound for the performance of randomized algorithms for V-MST-U which remains true even for instances with uniform query cost where the uncertainty graph is a cycle. Thus, the performance guarantee shown for \mysc{V-Random}$_C$ is best possible. Finally, we introduced the Minimum Spanning Star Problem under Explorable Uncertainty (MSS-U) and proved that no algorithm for MSS-U can achieve constant competitive ratio.

\noindent (V-)MST-U in itself is a problem which still deserves further investigation. A major open question in the setting of MST-U is whether there exists a randomized algorithm with a competitive ratio of $1.5$ for general instances as well or whether the lower bound of $1.5$ can be improved for instances where the graph is not necessarily a cactus. \\
\noindent As for V-MST-U, no randomized algorithm for general V-MST-U instances exists so far and even for the special case where cycles intersect in trivial vertices only, the performance guarantee of \mysc{V-Random}$_C$ relies on the fact that query costs are uniform. Note that no deterministic algorithm with constant competitive ratio for V-MST-U with non-uniform query costs is known either. \\
\noindent Moreover, different models of uncertainty exploration could be subject of further research. Consider for instance a scenario where installing a camera in a certain location allows to measure the distance between itself and all surrounding objects. A setting like this could motivate a hybrid model between edge and vertex uncertainty where edge weights are uncertain but known to lie inside given uncertainty sets and can be revealed upon querying an adjacent vertex.  

\newpage
\bibliography{main_article.bib}

\begin{thebibliography}{10}

\bibitem{borodin}
A.~Borodin and R.~El-Yaniv.
\newblock {\em Online Computation and Competitive Analysis}.
\newblock Cambridge University Press, 1998.

\bibitem{duerr}
C.~D{\"u}rr, T.~Erlebach, N.~Megow, and J.~Mei{\ss}ner.
\newblock {Scheduling with Explorable Uncertainty}.
\newblock In {\em 9th Innovations in Theoretical Computer Science Conference},
  volume~94 of {\em Leibniz International Proceedings in Informatics (LIPIcs)},
  pages 30:1--30:14. Schloss Dagstuhl--Leibniz-Zentrum fuer Informatik, 2018.

\bibitem{verification}
T.~Erlebach and M.~Hoffmann.
\newblock Minimum spanning tree verification under uncertainty.
\newblock In {\em Proceedings of the International Workshop on Graph-Theoretic
  Concepts in Computer Science}, pages 164--175, 2014.

\bibitem{erlebachhoffmann}
T.~Erlebach and M.~Hoffmann.
\newblock Query-competitive algorithms for computing with uncertainty.
\newblock {\em Bulletin of the European Association for Theoretical Computer
  Science}, 116, 2015.

\bibitem{erlebach}
T.~Erlebach, M.~Hoffmann, D.~Krizanc, M.~Mihal{\'{a}}k, and R.~Raman.
\newblock Computing minimum spanning trees with uncertainty.
\newblock {\em Proceedings of Symposium on Theoretical Aspects of Computer
  Science}, pages 277--288, 2008.

\bibitem{feder}
T.~Feder, R.~Motwani, L.~O'Callaghan, C.~Olston, and R.~Panigrahy.
\newblock Computing shortest paths with uncertainty.
\newblock {\em Journal of Algorithms}, 62(1):1–18, 2007.

\bibitem{focke}
J.~Focke, N.~Megow, and J.~Mei{\ss}ner.
\newblock {Minimum Spanning Tree under Explorable Uncertainty in Theory and
  Experiments}.
\newblock In {\em 16th International Symposium on Experimental Algorithms},
  volume~75 of {\em Leibniz International Proceedings in Informatics (LIPIcs)},
  pages 22:1--22:14. Schloss Dagstuhl--Leibniz-Zentrum fuer Informatik, 2017.

\bibitem{gargano}
L.~Gargano, M.~Hammar, P.~Hell, L.~Stacho, and U.~Vaccaroa.
\newblock Spanning spiders and light-splitting switches.
\newblock {\em Discrete Mathematics}, 285:83--95, 2004.

\bibitem{goerigk}
M.~Goerigk, M.~Gupta, J.~Ide, A.~Sch{\"o}bel, and S.~Sen.
\newblock The robust knapsack problem with queries.
\newblock {\em Computers and Operations Research}, 55:12--22, 2015.

\bibitem{kahan}
S.~Kahan.
\newblock A model for data in motion.
\newblock {\em 23rd Annual ACM Symposium on Theory of Computing (STOC’91)},
  pages 267--277, 1991.

\bibitem{megow}
N.~Megow, J.~Mei{\ss}ner, and M.~Skutella.
\newblock Randomization helps computing a minimum spanning tree under
  uncertainty.
\newblock {\em SIAM Journal on Computing}, 46(4):1217--1240, 2017.

\end{thebibliography}

\end{document}